\documentclass[10pt,twosides,a4paper]{book}
\usepackage[utf8]{inputenc}
\usepackage{bbm}
\usepackage{bm}
\usepackage{physics} 
\usepackage{amssymb}
\usepackage[pdftex]{graphicx}
\usepackage{latexsym}
\usepackage[small]{caption2}
\usepackage{amsmath}
\usepackage[english]{babel}
\usepackage{babel}
\usepackage{amsmath,amsthm}
\usepackage{rotating}
\usepackage{multirow}
\usepackage{graphicx}
\usepackage{tocloft}
\usepackage{bbm}
\usepackage{dsfont}
\usepackage{amsfonts}
\usepackage{accents}
\usepackage{pdfpages}
\usepackage[thinc]{esdiff}
\usepackage{hyperref}
\usepackage[numbers,sort&compress,square]{natbib}
\usepackage{natbib}
\usepackage{pdfsync}
\usepackage{fancyhdr}
\usepackage{scalerel,stackengine}
\usepackage{fancyhdr}
\usepackage{booktabs}
\usepackage{listings} 

\usepackage{algorithm}
\usepackage{algpseudocode}

\renewcommand{\cftchappresnum}{CHAPTER }
\newlength{\mylen}
\fancypagestyle{plain}{}

\usepackage{braket}

\usepackage{bbold}
\usepackage{mathtools}
\usepackage{mathrsfs} 

\input{tcilatex}  
\begin{document}

\chapter[Frank E. S. Steinhoff \newline
{\em Qutrit Clifford+T gates by two-body angular momentum couplings, rotations and one-axis-twistings}]{Qutrit Clifford+T gates by two-body angular momentum couplings, rotations and one-axis-twistings}
\label{chapter4}

\markboth{Qutrit Clifford+T gates by two-body angular momentum couplings}{F. E. S. Steinhoff}

{\large \textbf{Frank E. S. Steinhoff$^{1,2,a}$}}

\vspace{3mm}


\noindent {$^1$ Institute of Physics, University
of Brasilia, 70910-900, Brasilia, DF, Brazil}

\noindent {$^2$ International Center of Physics, Institute of Physics, University
of Brasilia, 70910-900, Brasilia, DF, Brazil}

\noindent {\texttt{$^{a}$frank.steinhoff@unb.br}}


\section{Abstract}
We develop an angular momentum representation and implementation of the Clifford+T set of unitaries for qutrits. We show that local gates from this set can be realized by the sole use of suitable rotations and one-axis-twisting operations, which are at most quadratic in the angular momentum operators and thus can be experimentally realized in many quantum systems. Controlled rotations are shown to only require linear angular momentum couplings and, as a consequence, the full qutrit Clifford+T set is shown to be expressed solely in terms of two-body angular momentum couplings, rotations and one-axis-twisting operations. By employing the Jordan-Schwinger map, we show an analogous implementation in terms of bosonic modes, improving on the number of modes with regard to a previous scheme. Moreover, we employ the cross-Kerr interaction in order to obtain any qutrit Clifford+T gate for bosonic modes. We illustrate our findings with simple schemes for preparing entangled states of interest.

\section{Introduction}

Higher dimensional systems have been identified as a worthwhile venue for quantum information processing \cite{frank.campbell,frank.nadish}, offering advantages when compared to their multi-qubit counterparts \cite{frank.nielsen}. The mathematical description of qudits - the $d$-dimensional analogs of qubits - is fairly well-understood when $d=p^n$, with $p$ prime \cite{frank.vourdas}, but the question remains on the feasibility of their actual physical realization. For example, a qudit density matrix demands $d^2-1$ observables in order to be reconstructed, and the numbers scale quickly with an increasing number of parties \cite{frank1}. 

In this vein, the author proposed in \cite{frank2} a broad framework for the representation and implementation of multi-qudit gates in terms of angular momentum interactions. Odd-dimensional systems are shown to demand significantly less $N$-body interactions in comparison to even-dimensional ones, suggesting a more detailed analysis of these systems. Since $d=3$ is the smallest non-trivial odd dimension, qutrits appear as a natural testbed of our formulation. 

In the original work \cite{frank2}, a quantum harmonic oscillator analog of the angular momentum representation is set forth, enabling the realization of arbitrary qudit gates in bosonic systems such as optical interferometers, superconducting circuits and Bose-Einstein condensates. Local qudit gates are implemented via linear operations, but, on the other hand, the number of bosonic modes required for each subsystem is $d$. In the present work we show that for qutrits we can reduce the number of modes required per subsystem to two; the trade-off is that this scheme requires some degree of nonlinearity for the expression of local unitaries in the form of Kerr interactions. The qutrit case is thus a very interesting special scenario, offering reasonable possibilities for the physical realization of quantum gates, be it in a angular momentum framework, where the interactions are at most quadratic in the operators, or in a two bosonic mode scheme, requiring Kerr nonlinearities. 

The work is organized as follows: in Section 1.3, we give a brief overview of the Clifford+T set of gates; in Section 1.4, the angular momentum framework is presented and it is shown that qutrit Clifford+T gates can be implemented via combinations of local rotations, one-axis-twisting operations and angular momentum linear couplings, which are typical interactions in many physical systems. In Section 1.5, we give a two bosonic mode analog of the angular momentum formulation, obtaining that qutrit Clifford+T gates can be implemented via appropriate combinations of linear and Kerr-nonlinear interactions between modes. In Section 1.6, we illustrate some possibilities of our findings in terms of the generation of certain entangled states of interest. In Section 1.7 we give our final considerations. For simplicity, in what follows we set $\hbar=1$. More details and references can be found in \cite{frank2}.

\section{Preliminaries}

In an arbitrary finite-dimensional system with orthonormal basis $\{|0\rangle, |1\rangle,\ldots,|d-1\rangle\}$, let us consider the gates 
\begin{eqnarray}
Z=\sum_{q=0}^{d-1}\omega^q|q\rangle\langle q|, \ \ \ X=\sum_{q=0}^{d-1}|q+1\rangle\langle q|, \ \ \omega=e^{2\pi i/d},
\end{eqnarray}
where operations are performed modulo $d$. The various products $X^jZ^k$ span a group called the {\it local Pauli group}; for example, for $d=2$ these products correspond to the usual Pauli matrices and the identity (modulo global phases). 

We can map $Z$ into $X$ via the finite-dimensional Fourier transform
\begin{eqnarray}
    F=d^{-1/2}\sum_{q,q'=0}^{d-1}\omega^{qq'}|q'\rangle\langle q|,
\end{eqnarray}
thus obtaining $X=FZF^{\dagger}$. More generally, when $d$ is a power of a prime number, we can map any element of the local Pauli group into any other by the action of unitaries from the {\it local Clifford group}. This is the group spanned by the unitaries
\begin{equation}
S(\xi,0,0)=\sum_{q=0}^{d-1}|\xi q\rangle\langle q|,\ S(1,\xi,0)=\sum_{q=0}^{d-1}\omega^{\xi q^2/2}|q\rangle\langle q|, \
S(1,0,\xi)=\sum_{q=0}^{d-1}\omega^{-\xi q^2/2}|p_q\rangle\langle p_q|, 
\end{equation}
where $|p_q\rangle=F|q\rangle$ are discrete versions of momentum eigenstates \cite{frank.vourdas}. Equivalently, within the full unitary group $U(d)$, the local Clifford group is the normalizer subgroup of the local Pauli group. 

In what follows, we restrict the discussion to the qutrit $d=3$ case. Let $\{|0_L\rangle,|1_L\rangle,|2_L\rangle\}$ denote the computational basis of the system considered. The elements of this basis can be any set of orthonormal states; for example, in a spin-$1$ system we can choose the $S_z$ eigenstates $|s=1,m=0\rangle,|s=1,m=\pm 1\rangle$, or the optical encoding $\{|1,0,0\rangle,|0,1,0\rangle,|0,0,1\rangle\}$ in \cite{frank2}.

Denoting by $[M]_c$ the matrix representation of the operator $M$ in the computational basis, we have  
\begin{eqnarray} 
[Z]_c=\left(\begin{array}{ccc}
{1}&{0}&{0}\\
{0}&{\omega}&{0}\\
{0}&{0}&{\omega^2}
\end{array}\right); \ \ \ [X]_c=\left(\begin{array}{ccc}
{0}&{0}&{1}\\
{1}&{0}&{0}\\
{0}&{1}&{0}
\end{array}\right) 
\end{eqnarray}
and the qutrit Fourier transform is given by
\begin{eqnarray}
    [F]_c=\frac{1}{\sqrt{3}}\left(\begin{array}{ccc}
{1}&{1}&{1}\\
{1}&{\omega}&{\omega^2}\\
{1}&{\omega^2}&{\omega}
\end{array}\right).
\end{eqnarray}
We can construct the qutrit momentum basis by applying the Fourier operator to the computational basis elements:
\begin{eqnarray}
    |+\rangle &\equiv& |p_0\rangle=F|0_L\rangle=\frac{1}{\sqrt{3}}(|0_L\rangle+|1_L\rangle+|2_L\rangle),\\
    |\omega\rangle &\equiv& |p_1\rangle=F|1_L\rangle=\frac{1}{\sqrt{3}}(|0_L\rangle+\omega|1_L\rangle+\omega^2|2_L\rangle),\\
    |\omega^2\rangle &\equiv& |p_2\rangle=F|2_L\rangle=\frac{1}{\sqrt{3}}(|0_L\rangle+\omega^2|1_L\rangle+\omega|2_L\rangle).
\end{eqnarray}
A convenient set of Clifford operators are the permutations:
\begin{equation}
[X_{12}]_c=\left(\begin{array}{ccc}
{1}&{0}&{0}\\
{0}&{0}&{1}\\
{0}&{1}&{0}
\end{array}\right), \ 
[X_{02}]_c=\left(\begin{array}{ccc}
{0}&{0}&{1}\\
{0}&{1}&{0}\\
{1}&{0}&{0}
\end{array}\right), \ [X_{01}]_c=\left(\begin{array}{ccc}
{0}&{1}&{0}\\
{1}&{0}&{0}\\
{0}&{0}&{1}
\end{array}\right).
\end{equation}  

For two-qutrits, the global computational basis is given by the tensor product of the local computational bases in lexicographical order: 
\begin{equation}
    \{|0_L,0_L\rangle,|0_L,1_L\rangle,|0_L,2_L\rangle,|1_L,0_L\rangle,|1_L,1_L\rangle,|1_L,2_L\rangle,|2_L,0_L\rangle,|2_L,1_L\rangle,|2_L,2_L\rangle\}
\end{equation}

Given $W$ unitary, a {\it controlled gate} is a two-qutrit unitary of the form 
\begin{eqnarray}
    CW=|0_L\rangle\langle 0_L|\otimes I+|1_L\rangle\langle 1_L|\otimes W+|2_L\rangle\langle 2_L|\otimes W^2.
\end{eqnarray}
In the computational basis, a controlled gate takes a block diagonal form
\begin{eqnarray}
    [CW]_c=\left(\begin{array}{c|c|c}
{I}&{}&{}\\
\hline{}&{W}&{}\\
\hline{}&{}&{W^2}
\end{array}\right).
\end{eqnarray}
The controlled gates constructed from the $Z$ and $X$ gates, i.e., $CZ$ and $CX$, together with the gates from the local Clifford groups of each subsystem, span the {\it Clifford group} on $n-$qutrits. This group constitutes a fault-tolerant gate set, enabling the design of quantum error correcting codes. Interestingly, this gate set can be efficiently simulated by a probabilistic classical computation scheme, a result known as the {\it Gottesmann-Knill theorem} \cite{frank.gkt}. To move beyond this limitation, we need the assistance of non-Clifford gates, such as multi-controlled gates, or the so-called ``magic" $T$ gate \cite{frank.3T} given by
\begin{eqnarray}
[T]_c=\left(\begin{array}{ccc}
{1}&{0}&{0}\\
{0}&{\eta}&{0}\\
{0}&{0}&{\eta^{-1}}
\end{array}\right)
\end{eqnarray}
where $\eta=e^{2\pi i/9}$. Fortunately, the qutrit Clifford+T set is approximately universal, as shown in \cite{frank.3univ} and we can build with it arbitrary sets of qutrit multi-controlled Clifford unitaries \cite{frank.3gates}.

\section{Angular momentum realization}

\subsection{Local gates}

As is well-known \cite{frank.sakurai}, we can represent a finite-dimensional system via angular momentum operators $J_x$, $J_y$ and $J_z$ that span the Lie algebra $su(2)$ via the commutation relations 
\begin{eqnarray}
[J_x,J_y]=iJ_z, \ \ \ [J_y,J_z]=iJ_x ,\ \ \ [J_z,J_x]=iJ_y.
\end{eqnarray}
A basis of the state space (in each irreducible representation) is built by taking the $d=2j+1$ simultaneous eigenstates $|j;m\rangle$ of the commuting operators $J_z$ and $J^2=J^2_x+J^2_y+J^2_z$, with the relations $J^2|j;m\rangle=j(j+1)|j;m\rangle$, $J_z|j;m\rangle=m|j;m\rangle$; here $j=0,1/2,1,3/2,\ldots$ and $m=-j,-j+1,\ldots,j-1,j$ are the possible values of the quantum numbers. 

Since we are interested in qutrits, we restrict our discussion to the $j=1$ case. For convenience, we drop the quantum number $j$ and indicate the angular momentum basis by $\{|m=1\rangle,|m=0\rangle,|m=-1\rangle\}$. In this ordered basis, a given operator $M$ has matrix representation $[M]_j$; for example, we have
\begin{equation}
[J_x]_j= \frac{1}{\sqrt{2}}\left(\begin{array}{ccc}
{0}&{1}&{0}\\
{1}&{0}&{1}\\
{0}&{1}&{0}
\end{array}\right); \ 
[J_y]_j= \frac{1}{\sqrt{2}}\left(\begin{array}{ccc}
{0}&{-i}&{0}\\
{i}&{0}&{-i}\\
{0}&{i}&{0}
\end{array}\right); \ 
[J_z]_j= \left(\begin{array}{ccc}
{1}&{0}&{0}\\
{0}&{0}&{0}\\
{0}&{0}&{-1}
\end{array}\right). 
\end{equation}
A difference to the $j=1/2$ case is the presence of quadratic operators such as
\begin{equation}
[J_x^2]_j= \frac{1}{2}\left(\begin{array}{ccc}
{1}&{0}&{1}\\
{0}&{2}&{0}\\
{1}&{0}&{1}
\end{array}\right); \ 
[J^2_y]_j= \frac{1}{2}\left(\begin{array}{ccc}
{1}&{0}&{-1}\\
{0}&{2}&{0}\\
{-1}&{0}&{1}
\end{array}\right); \
[J^2_z]_j= \left(\begin{array}{ccc}
{1}&{0}&{0}\\
{0}&{0}&{0}\\
{0}&{0}&{1}
\end{array}\right). 
\end{equation}
But since $J^3_l=J_l$, $l=x,y,z$, we get simplified expressions for unitaries of the form
\begin{eqnarray}
R(l,\phi)&=&\exp(i\phi J_l)=I+i\sin\phi J_l+(\cos\phi-1)J_l^2,\\
U_{oat}(l,\phi)&=&\exp(i\phi J_l^2)=I+(e^{i\phi}-1)J^2_l, \ \ \ l=x,y,z.
\end{eqnarray} 
The unitaries $R(l,\phi)$ represent rotations on an axis $l$ by an angle $\phi$; in spin systems, these can be implemented via the action of a uniform external magnetic field. A unitary $U_{oat}(l,\phi)$ corresponds to the so-called one-axis-twisting (OAT) operation by $\phi$ with respect to an axis $l$, being generated by, e.g., quadrupolar magnetic interactions and have an important role in the preparation of spin-squeezed states \cite{frank.spinsq}. 

    Rotations about the $z$-axis are diagonal and read as
\begin{eqnarray}
    [R(z,\phi)]_j=[\exp(i\phi J_z)]_j=
\left(\begin{array}{ccc}
{e^{i\phi}}&{0}&{0}\\
{0}&{1}&{0}\\
{0}&{0}&{e^{-i\phi}}
\end{array}\right)
\end{eqnarray}
and, particularly for $\phi=2\pi/3$, we have
\begin{eqnarray}
    [R(z,2\pi/3)]_j=
\left(\begin{array}{ccc}
{e^{2\pi i/3}}&{0}&{0}\\
{0}&{1}&{0}\\
{0}&{0}&{e^{-2\pi i/3}}
\end{array}\right)=
\left(\begin{array}{ccc}
{\omega}&{0}&{0}\\
{0}&{1}&{0}\\
{0}&{0}&{\omega^2}
\end{array}\right),
\end{eqnarray}
given that $\omega^{-1}=\omega^2$. This suggests the encoding $|0_L\rangle\equiv|m=0\rangle$, $|1_L\rangle\equiv|m=1\rangle$ and $|2_L\rangle\equiv|m=-1\rangle$. The computational basis $\{|0_L\rangle,|1_L\rangle,|2_L\rangle\}$ amounts then to a reordering of the angular momentum basis and we obtain $[R(z,2\pi/3)]_c=[Z]_c$, i.e., $Z=\exp[(2\pi i/3) J_z]$ in this convention. 
An important permutation gate is given by the $x$-rotation
\begin{eqnarray}
[R(x,\pi)]_j=\left(\begin{array}{ccc}
{1}&{0}&{0}\\
{0}&{1}&{0}\\
{0}&{0}&{1}
\end{array}\right)-\left(\begin{array}{ccc}
{1}&{0}&{1}\\
{0}&{2}&{0}\\
{1}&{0}&{1}
\end{array}\right)=\left(\begin{array}{ccc}
{0}&{0}&{-1}\\
{0}&{-1}&{0}\\
{-1}&{0}&{0}
\end{array}\right),
\end{eqnarray}
which under the computational reordering is $[R(x,\pi)]_c=-[X_{12}]_c$.

In order to obtain the full local Pauli group, we need the $X=FZF^{\dagger}=\exp[(4\pi i/3)\Theta_z]$ gate\footnote{There is some global phase (gauge) freedom here that the reader should pay attention to.}, where
\begin{eqnarray}
[\Theta_z]_j= [F]_j[J_z]_j[F^{\dagger}]_j=\frac{1}{\sqrt{3}}\left(\begin{array}{ccc}
{0}&{i}&{-i}\\
{-i}&{0}&{i}\\
{i}&{-i}&{0}
\end{array}\right)
\end{eqnarray} 
is the so-called {\it Pegg-Barnett hermitian phase operator} \cite{frank.pb1,frank.pb2,frank.pb3} in $d=3$. Straightforward calculations show that
\begin{eqnarray}
\Theta_z= \sqrt{\frac{1}{3}}\{J_y,J_x\}-\sqrt{\frac{2}{3}}J_y. \label{frank.3pgo}
\end{eqnarray} 
Alternatively, $\Theta_z=\cos\alpha\{J_y,J_x\}-\sin\alpha J_y$, where $\alpha=\tan^{-1}(\sqrt{2})$ is a constant known as the {\it magic angle} \cite{frank.ma}, appearing in areas such as nuclear magnetic resonance and spectroscopy. With this expression, we can deduce a manner of obtaining $\Theta_z$ from $J_z$ via suitable rotations and OAT gates:
\begin{eqnarray}
    \Theta_z=[U_{oat}(y,-\pi/2)R(x,-\alpha)]J_z[R(x,\alpha)U_{oat}(y,\pi/2)].
\end{eqnarray}
From this discussion, we obtain as well an expression for the qutrit Fourier transform
\begin{equation}
    F=-U_{oat}(y,-\pi/2)R_x(-\alpha)e^{-i(\pi/2)\Pi_0}=U_{oat}(y,-\pi/2)R_x(-\alpha)U_{oat}(z,-\pi/2)e^{i\pi/2}
\end{equation}
where $\Pi_0=|0_L\rangle\langle 0_L|=I-J_z^2$; hence, we can realize the operators $F$, $\Theta_z$ and $X$ in angular momentum systems by convenient OAT and rotation gates. This is also true for the remaining Clifford gates
\begin{eqnarray}
S(1,1,0)=\exp[(4\pi i/3) J_z^2]; \ \ S(1,2,0)=\exp[(2\pi i/3) J_z^2]; \\ S(1,0,1)=\exp[(2\pi i/3) \Theta_z^2]; \ \ S(1,0,2)=\exp[(4\pi i/3) \Theta_z^2],  
\end{eqnarray}
given that $\Theta_z^2=FJ_z^2F^{\dagger}$. An interesting alternative expression is $\Theta_z^2=-\sqrt{2}J_x+2J_y^2+J_z^2$, which corresponds to the Lipkin-Meshkov-Glick interaction \cite{frank.lmg1,frank.lmg2}. 
Surprisingly, the qutrit $T$ gate is a simple $z$-rotation $T=\exp[(2\pi i/9) J_z]=R(z,2\pi/9)$ and is thus a less expensive resource in our formulation.

\subsection{Controlled gates}

Having addressed the implementation of local gates, we now consider the expressions for the $CZ$ and $CX$ gates. We start from    
\begin{eqnarray}
[J^a_z\otimes J_z^b]_j&=&\left(\begin{array}{ccc}
{1}&{0}&{0}\\
{0}&{0}&{0}\\
{0}&{0}&{-1}
\end{array}\right)\otimes\left(\begin{array}{ccc}
{1}&{0}&{0}\\
{0}&{0}&{0}\\
{0}&{0}&{-1}
\end{array}\right)\\
&=&\left(
\begin{array}{ccc|ccc|ccc}
           {1}&{}&{}&{}&{}&{}&{}&{}&{}  \\
           {}&{0}&{}&{}&{}&{}&{}&{}&{}  \\
           {}&{}&{-1}&{}&{}&{}&{}&{}&{}  \\
    \hline {}&{}&{}&{0}&{}&{}&{}&{}&{}  \\
           {}&{}&{}&{}&{0}&{}&{}&{}&{}  \\
           {}&{}&{}&{}&{}&{0}&{}&{}&{}  \\
   \hline  {}&{}&{}&{}&{}&{}&{-1}&{}&{}  \\
           {}&{}&{}&{}&{}&{}&{}&{0}&{}  \\
           {}&{}&{}&{}&{}&{}&{}&{}&{1}   
\end{array}\right).
\end{eqnarray}
This is a typical interaction in systems described by angular momentum. For example, interactions between spin dipoles $\mathbf{S}_a$ and $\mathbf{S}_b$ are usually given by $\mathbf{S}_a\cdot\mathbf{S}_b$ and spin-orbit interactions take the form $\mathbf{S}\cdot\mathbf{L}$. The unitaries resulting from such interactions are then given by  
\begin{eqnarray*}
    \left[\exp[i\phi J^a_z\otimes J^b_z]\right]_j=\left(
\begin{array}{ccc|ccc|ccc}
           {e^{i\phi}}&{}&{}&{}&{}&{}&{}&{}&{}  \\
           {}&{1}&{}&{}&{}&{}&{}&{}&{}  \\
           {}&{}&{e^{-i\phi}}&{}&{}&{}&{}&{}&{}  \\
    \hline {}&{}&{}&{1}&{}&{}&{}&{}&{}  \\
           {}&{}&{}&{}&{1}&{}&{}&{}&{}  \\
           {}&{}&{}&{}&{}&{1}&{}&{}&{}  \\
   \hline  {}&{}&{}&{}&{}&{}&{e^{-i\phi}}&{}&{}  \\
           {}&{}&{}&{}&{}&{}&{}&{1}&{}  \\
           {}&{}&{}&{}&{}&{}&{}&{}&{e^{i\phi}}   
\end{array}\right)=[CR(z,\phi)]_j
\end{eqnarray*}
and setting $\phi=2\pi/3$, we obtain
\begin{eqnarray*}
   \left[\exp[(2\pi i/3) J^a_z\otimes J^b_z]\right]_j=\left(
\begin{array}{ccc|ccc|ccc}
           {\omega}&{}&{}&{}&{}&{}&{}&{}&{}  \\
           {}&{1}&{}&{}&{}&{}&{}&{}&{}  \\
           {}&{}&{\omega^2}&{}&{}&{}&{}&{}&{}  \\
    \hline {}&{}&{}&{1}&{}&{}&{}&{}&{}  \\
           {}&{}&{}&{}&{1}&{}&{}&{}&{}  \\
           {}&{}&{}&{}&{}&{1}&{}&{}&{}  \\
   \hline  {}&{}&{}&{}&{}&{}&{\omega^2}&{}&{}  \\
           {}&{}&{}&{}&{}&{}&{}&{1}&{}  \\
           {}&{}&{}&{}&{}&{}&{}&{}&{\omega}   
\end{array}\right).
\end{eqnarray*}
The computational reordering results in
\begin{eqnarray*}
    \left[ \exp[(2\pi i/3) J^a_z\otimes J^b_z]\right]_c=\left(
\begin{array}{ccc|ccc|ccc}
           {1}&{}&{}&{}&{}&{}&{}&{}&{}  \\
           {}&{1}&{}&{}&{}&{}&{}&{}&{}  \\
           {}&{}&{1}&{}&{}&{}&{}&{}&{}  \\
    \hline {}&{}&{}&{1}&{}&{}&{}&{}&{}  \\
           {}&{}&{}&{}&{\omega}&{}&{}&{}&{}  \\
           {}&{}&{}&{}&{}&{\omega^2}&{}&{}&{}  \\
   \hline  {}&{}&{}&{}&{}&{}&{1}&{}&{}  \\
           {}&{}&{}&{}&{}&{}&{}&{\omega^2}&{}  \\
           {}&{}&{}&{}&{}&{}&{}&{}&{\omega}   
\end{array}\right)=[CZ]_c
\end{eqnarray*}
and we conclude that $CZ=\exp[(2\pi i/3) J^a_z\otimes J^b_z]$. The $CX$ gate can be obtained through $CX=\exp[(4\pi i/3) J^a_z\otimes \Theta^b_z]$, where $\Theta^b_z$ is obtained from $J_z^b$ by the local rotations and OAT in (\ref{frank.3pgo}), or alternatively via $CX=(I\otimes F)CZ(I\otimes F^{\dagger})$; the Fourier transform, however, require an extra OAT gate, due to the projection onto $|m=0\rangle\langle m=0|$.

\section{Quantum harmonic oscillator realizations}

\subsection{The Jordan-Schwinger map}

We can map the angular momentum operators into two bosonic modes via 
\begin{eqnarray}
J_x=\frac{1}{2}(a^{\dagger}b+ab^{\dagger}), \ \ \ J_y=\frac{1}{2i}(a^{\dagger}b-ab^{\dagger}), \ \ \ J_z=\frac{1}{2}(a^{\dagger}a-b^{\dagger}b), \label{frank.jsm}
\end{eqnarray}
which is known as the {\it Jordan-Schwinger map} \cite{frank.jsm,frank.yurke,frank.puri,frank.su2su11}. Here the operators $a$ and $b$ satisfy canonical commutation relations $[a,a^{\dagger}]=I=[b,b^{\dagger}]$ and $[a,b]=[a,b^{\dagger}]=[a^{\dagger},b]=[a^{\dagger},b^{\dagger}]=0$. This isomorphism between Lie algebras allows us to give physical realizations of the angular momentum operators in a plethora of bosonic systems such as optical interferometers, double-well Bose-Einstein condensates, circuit QED and more. In these systems, the operators (\ref{frank.jsm}) represent typical interaction terms; for example, in optical systems $J_x$ and $J_y$ correspond to beam-splitting interactions \cite{frank.scully}, while in double-well Bose-Einstein condensates these operators represent tunneling between the modes.  

In order to relate the two representations, we observe that $J^2=\frac{N}{2}(\frac{N}{2}+I)$ and $J_z=\frac{N_a-N_b}{2}$, where $N_a=a^{\dagger}a$, $N_b=b^{\dagger}b$ and $N=N_a+N_b$ are the respective local and total number operators. Let $N_a|n_a\rangle=n_a|n_a\rangle$, $N_b|n_b\rangle=n_b|n_b\rangle$ and $N|n\rangle=n|n\rangle$, where obviously $n=n_a+n_b$. Then we have the following relation between the quantum numbers of each representation: $j=(n_a+n_b)/2$, $m=(n_a-n_b)/2$ and we can write $|n_a,n_b\rangle\equiv|j+m,j-m\rangle$. For the qutrit case $j=1$, this suggests the encoding
\begin{eqnarray}
    |0_L\rangle&\equiv&|j=1;m=0\rangle\equiv|n_a=1,n_b=1\rangle,\\ 
    |1_L\rangle&\equiv&|j=1;m=+1\rangle\equiv|n_a=2,n_b=0\rangle,\\
    |2_L\rangle&\equiv&|j=1;m=-1\rangle\equiv|n_a=0,n_b=2\rangle,
\end{eqnarray}
where the total number $n=2$ is fixed. The rotations obtained from exponentiation of the operators (\ref{frank.jsm}) correspond to mode splitting transformations of the modes $a$ and $b$ in the case of $x$ and $y$ rotations, or phase-shifting unitaries in the case of $z$ rotations \cite{frank.yurke}.

\subsection{Kerr interactions}

In order to implement more complex gates, we need interactions that are quadratic in $N_a$ and $N_b$. These are known as {\it Kerr nonlinearities} \cite{frank.boyd} and they appear in basically two types. The first type is a one-mode interaction of the form
\begin{eqnarray}
    H^{(a)}_{sk}=\chi (a^{\dagger}a)^2=\chi N_a^2,
\end{eqnarray}
known as {\it self-Kerr interaction} and the second type is a two-mode coupling between the number operators
\begin{eqnarray}
    H^{(a,b)}_{ck}=\chi' a^{\dagger}ab^{\dagger}b=\chi'N_aN_b,
\end{eqnarray}
called {\it cross-Kerr interaction}; the constants $\chi$ and $\chi'$ correspond to the respective interaction strengths. In optical setups these values tend to be very weak, but in superconducting circuits these interactions are usually tunable and can acquire strong values. On the other hand, the latter suffer from scalability issues, whereas modern approaches have been developed to address the limitations of the former \cite{frank.ck,frank.sk,frank.sqkerr}.

\subsection{OAT via self-Kerr nonlinearities}

As discussed previously, most local Clifford gates are implemented via suitable rotations and OAT unitaries. We describe here the procedure outlined originally in \cite{frank.spinsq} for the implementation of arbitrary OAT operations. Starting from the simple relation
\begin{eqnarray}
    N^2_a+N^2_b=\frac{(N_a+N_b)^2}{2}+\frac{(N_a-N_b)^2}{2}=\frac{N^2}{2}+2J^2_z,
\end{eqnarray}
we obtain
\begin{eqnarray}
     J_z^2=  \frac{N^2_a}{2}+\frac{N^2_b}{2}-\frac{N^2}{4}.
\end{eqnarray}
A $z$-axis OAT then reads
\begin{eqnarray}
 U_{oat}(z,\phi)&=&\exp[i\phi J_z^2]=\exp\left[i\phi\left(\frac{N^2_a}{2}+\frac{N^2_b}{2}-\frac{N^2}{4}\right)\right]   \\
 &=& \exp\left[\frac{i\phi}{\chi}H^{(a)}_{sk}\right]\exp\left[\frac{i\phi}{\chi}H^{(b)}_{sk}\right]e^{-i\phi n^2}
\end{eqnarray}
where we assume that the total number has a fixed value $n$. We conclude that a $z$-axis OAT can be implemented by two independent self-Kerr interactions of appropriate strengths. Arbitrary OAT can then be implemented by the action of suitable rotations on $ U_{oat}(z,\phi)$, corresponding to a self-Kerr augmented Mach-Zender interferometer. As shown previously, local Clifford gates are finite sequences of OAT and rotations and hence can be realized in two modes by suitable combinations of self-Kerr interactions and linear couplings between the modes.

\subsection{Implementations via cross-Kerr nonlinearities}

The remaining gates can be implemented using cross-Kerr interactions. For example, the projector $\Pi_0=|0_L\rangle\langle 0_L|\equiv |m=0\rangle\langle m=0|$, required for the implementation of the qutrit Fourier, is given by
\begin{eqnarray}
    |m=0\rangle\langle m=0|=\frac{N^2}{4}-J^2_z=N_aN_b
\end{eqnarray}
Indeed, we can check directly that $N_aN_b|2,0\rangle=N_aN_b|0,2\rangle=0,N_aN_b|1,1\rangle=|1,1\rangle$, and the cross-Kerr coupling acts as a projector\footnote{Differently from \cite{frank2}, the presence of quadratic terms in $N_a$ and $N_b$ for the computational basis projectors hinders the direct implementation of hard-controlled gates and we need the more costly alternative in \cite{frank.3gates}, Eq. (9). } onto $|1,1\rangle\langle 1,1|$. We stress that this is valid only for the fixed $n=2$ case. 

The cross-Kerr coupling also allows us to build a $CZ$ gate between two arbitrary subsystems $\mu$ and $\nu$; observing that
\begin{eqnarray}
    J_z^{(\mu)}J_z^{(\nu)}&=&\left(\frac{a^{\dagger}_{\mu}a_{\mu}-b^{\dagger}_{\mu}b_{\mu}}{2}\right)\left(\frac{a^{\dagger}_{\nu}a_{\nu}-b^{\dagger}_{\nu}b_{\nu}}{2}\right) \\
    &=&\frac{1}{4}\left(a^{\dagger}_{\mu}a_{\mu}a^{\dagger}_{\nu}a_{\nu}-a^{\dagger}_{\mu}a_{\mu}b^{\dagger}_{\nu}b_{\nu}-b^{\dagger}_{\mu}b_{\mu}a^{\dagger}_{\nu}a_{\nu}+b^{\dagger}_{\mu}b_{\mu}b^{\dagger}_{\nu}b_{\nu}\right)\\
    &=&\frac{1}{4\chi'}\left(H^{(a_{\mu}a_{\nu})}_{ck}-H^{(a_{\mu}b_{\nu})}_{ck}-H^{(b_{\mu}a_{\nu})}_{ck}+H^{(b_{\mu}b_{\nu})}_{ck}\right),
\end{eqnarray}
we obtain that the $CZ$ gate can be implemented by a sequence of four cross-Kerr interactions:
\begin{eqnarray*}
    CZ&=&\exp\left[\frac{2\pi i}{3}J_z^{(\mu)}J_z^{(\nu)}\right]=\exp\left[\frac{\pi i}{6\chi'}\left(H^{(a_{\mu}a_{\nu})}_{ck}-H^{(a_{\mu}b_{\nu})}_{ck}-H^{(b_{\mu}a_{\nu})}_{ck}+H^{(b_{\mu}b_{\nu})}_{ck}\right)\right]\\
    &=& \exp\left[\frac{\pi i}{6\chi'}H^{(a_{\mu}a_{\nu})}_{ck}\right]\exp\left[\frac{11\pi i}{6\chi'}H^{(a_{\mu}b_{\nu})}_{ck}\right]\exp\left[\frac{11\pi i}{6\chi'}H^{(b_{\mu}a_{\nu})}_{ck}\right]\exp\left[\frac{\pi i}{6\chi'}H^{(b_{\mu}b_{\nu})}_{ck}\right].
\end{eqnarray*}
The $CX$ gate can be obtained from the $CZ$ above by applying the local gates that bring $J_z^{(\nu)}$ into $\Theta_z^{(\nu)}$, for example. Hence, we can implement any qutrit Clifford+T gate via suitable linear couplings and Kerr nonlinearities. 

\section{Application: entangled states preparation}

\subsection{Maximally entangled state between two bosonic modes}

As a simple and straightforward application of our findings, we have the preparation of a maximally entangled state between two modes:
\begin{eqnarray*}
    |+\rangle = F|0_L\rangle=\frac{1}{\sqrt{3}}(|0_L\rangle+|1_L\rangle+|2_L\rangle)\equiv\frac{1}{\sqrt{3}}(|2,0\rangle+|1,1\rangle+|0,2\rangle)
\end{eqnarray*}
We are not aware of a similar (deterministic) proposal in the literature for the preparation of this type of state. In the angular momentum representation, this is a highly non-classical superposition\footnote{The nomenclature is somewhat ambiguous: the state has maximal coherence (superposition), being an example of a non-$SU(2)$ coherent state.} $(1/\sqrt{3})(|m=1\rangle+|m=0\rangle+|m=-1\rangle)$. Moreover, we can draw parallels with the Hong-Ou-Mandell scenario \cite{frank.hom}, where the state produced is $(1/\sqrt{2})(|0,2\rangle-|2,0\rangle)$, coming from sending $|1,1\rangle$ through a balanced beam-splitter. 

\subsection{Qutrit graph states}

Given a multi-graph $G=(V,E)$, we can define a genuinely multipartite entangled state $|G\rangle$ via the expression 
\begin{eqnarray}
    |G\rangle=\prod_{e\in E}(CZ_e)^{g_e}|+\rangle^V
\end{eqnarray}
where $g_e\in\{0,1,2\}$ and $|+\rangle^V\equiv\bigotimes_{v\in V}|+_v\rangle$. States in this class are called {\it qudit graph states} \cite{frank.gs} and are directly connected to the qudit Clifford group.  

Since we know now how to realize qutrit Clifford gates both via angular momentum interactions as well as by interactions of two bosonic modes, we can in principle prepare any qutrit graph state. Let us consider the simplest example of a qutrit graph state $|G_{GHZ}\rangle$ that corresponds to the qutrit GHZ state: 
\begin{equation}
    |G_{GHZ}\rangle=CZ_{12}CZ_{13}|+_1,+_2,+_3\rangle=\frac{1}{\sqrt{3}}\left[|0_1,+_2,+_3\rangle+|1_1,\omega_2,\omega_3\rangle+|2_1,\omega^2_2,\omega^2_3\rangle\right].
\end{equation}
The GHZ graph here is $G_{GHZ}=(E=\{\{1,2\},\{1,3\}\},V=\{1,2,3\})$. Interestingly, acting locally on the subsystem $1$ with the Clifford gate $X_{12}$ amounts to changing both edges multiplicities from $g_e=1$ to $2$. The standard GHZ state can be recovered by local Fourier unitaries:
\begin{eqnarray}
    |GHZ\rangle=F_2^{\dagger}F_3^{\dagger}|G_{GHZ}\rangle=\frac{1}{\sqrt{3}}\left[|0_1,0_2,0_3\rangle+|1_1,1_2,1_3\rangle+|2_1,2_2,2_3\rangle\right].
\end{eqnarray}

\subsection{$j=1$ angular momentum graph states}

In \cite{frank2}, a class of states known as {\it angular momentum hypergraph states} was defined, being a variation of the set of {\it qudit hypergraph states} \cite{frank3} more directly related to the angular momentum representation. A special subset is found in the so-called {\it angular momentum graph states}, where from a weighted graph $G=(V,E)$ we associate a state
\begin{eqnarray}
    |J_G\rangle=\prod_{e\in E}\exp\left[i\phi_eJ_z^{(e)}\right]|x+\rangle^V=\exp\left[i\sum_{e\in E}\phi_eJ_z^{(e)}\right]|x+\rangle^V;
\end{eqnarray}
here $J_z^{(e)}=J_z^{\mu}J_z^{\nu}$, with $e=\{\mu,\nu\}$, the vertex state $|x+\rangle$ is the eigenstate of $J_x$ with maximal eigenvalue and the edge weights $\phi_e$ can assume any real value. This relaxation on the weights allow us to consider interactions of any strength; in the optical case, in the spirit of \cite{frank.wgs}, we can now embrace the weak nature of Kerr nonlinearities in order to design relevant gates and states. 
    
\begin{figure}[tbh]
\begin{center}
\includegraphics[width=0.99\textwidth]{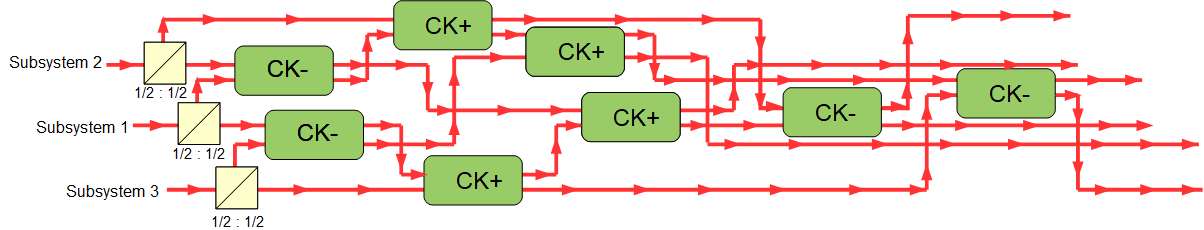} {}
\end{center}
\caption{Interferometric diagram for the preparation of a qutrit angular momentum graph state $|J_{GHZ}\rangle$. The input state in each subsystem is the two-photon state $|2,0\rangle$.}
\label{frank.fig1}
\end{figure}

For $j=1$ each vertex state has the expression
\begin{eqnarray}
    |x+\rangle=\frac{1}{2}(|m=1\rangle+\sqrt{2}|m=0\rangle+|m=-1\rangle).
\end{eqnarray}
A GHZ-like state in this class is then given by
\begin{eqnarray}
    |J_{GHZ}\rangle&=&\exp\left[i\phi(J_z^{(1)}J_z^{(2)}+J_z^{(1)}J_z^{(3)})\right] |x_1+,x_2+,x_3+\rangle\\
    &=&CR_{12}(z,\phi)CR_{13}(z,\phi)|x_1+,x_2+,x_3+\rangle\\ 
    &=&\frac{1}{2}\left[|1_1,\phi_2,\phi_3\rangle+\sqrt{2}|0_1,x_2+,x_3+\rangle+|-1_1,-\phi_2,-\phi_3\rangle\right];
\end{eqnarray}
where $|\phi\rangle=R(z,\phi)|x+\rangle$. The state has Schmidt rank $3$ (in the multipartite sense) when $\phi$ is not an integer multiple of $\pi$, $2$ when it is an odd multiple of $\pi$ and obviously $1$ (separable) when it is an even multiple of $\pi$. When $\phi=\pm2\pi/3$, we can show that the state is equivalent under SLOCC to the GHZ state. A possible optical scheme to generate this state is illustrated in (\ref{frank.fig1}).

Since Kerr interactions do not alter local or total number values, we believe techniques from \cite{frank4} may be applicable to the nonlinear interferometers described here. 

\section{Conclusions}
In the present work we obtained physical realizations of any qutrit Clifford+T gates in two scenarios: an angular momentum one and its two bosonic mode analog. In the first framework, we showed that these gates can be implemented via sequences of rotations, OAT unitaries and two-body couplings. In the second, the corresponding gate decompositions are made in terms of linear operations and Kerr nonlinearities. We illustrated our results with simple schemes to implement entangled states, with special attention to the set of $j=1$ angular momentum graph states, which can be experimentally realized with experimental techniques currently available. 

We considered an idealized lossless scenario, since a full analysis of the many sources of errors go beyond the scope of the present work. Similarly, we did not consider the optimization of circuits or of the experimental setups. We believe that in its current form, our scheme is suited for a few-body regime, where the complexity of operations can be reasonably controlled. However, we also believe that our formulation can be useful in the analysis of certain many-body quantum networks governed by interactions similar to the ones described here. Moreover, as the original work \cite{frank2} suggests, $N$-body angular momentum interactions - or their Kerr analogs - can significantly reduce the number of physical resources required to implement multipartite gates and states. Hence, their theoretical and experimental characterization is a worthwhile pursuit.

\section*{Acknowledgments}
The author is thankful to Daniel J. Brod, Alexandre Ribeiro and Eduardo I. Duzzioni for ideas and discussions. The author also acknowledges financial support from the Physics Institute of University of Bras\'ilia, grant 170136.1050A000AP.154228.\\MGY01N0104N, as well as partial funding from the brazilian agency FINATEC.

\renewcommand{\bibname}{References} \begingroup
\let\cleardoublepage\relax

\endgroup

\end{document}